# A Kinetic Model of Solar Wind Generation and Heating by Kinetic Alfvén Wave Turbulence

Philip A. Isenberg and Bernard J. Vasquez

*Space Science Center and Department of Physics and Astronomy, University of New Hampshire, Durham NH, 03824, USA. Corresponding author: phil.isenberg@unh.edu*

**Abstract.** We present results from a kinetic model of collisionless gyrotropic coronal hole protons heated by cyclotron and Landau resonant dissipation of critically balanced kinetic Alfvén waves. The model incorporates the kinetic effects of gravity, ambipolar electric field, ponderomotive force of the large-scale Alfvén waves and the mirror force in a super-radially expanding flux tube. The flow speed is self-consistently obtained as the bulk flow of the proton distribution. Two cases, taking the intensities of the turbulent spectra to be balanced in the parallel propagation direction or imbalanced at a ratio of 9:1 show almost no difference. The distributions develop parallel extensions outward due to the Landau resonance, but exhibit very little perpendicular heating under the parameter choices used here. The speeds and temperatures fall well short of the requirements for a fast solar wind. Rather than continuing to explore the parameter space of this system, we propose that a modified model using turbulent spectra based on the recent helicity barrier scenario would be an appropriate next step.

## INTRODUCTION

One of the most stubborn problems remaining in solar wind studies is the identification of the plasma processes responsible for the generation of the fast solar wind in the corona. A commonly invoked scenario is that a sufficient flux of large-scale shear Alfvén wave power is produced at low altitudes by some combination of gradual and explosive plasma motions. These waves propagate outward and initiate a turbulent cascade. The turbulence then transports the fluctuation energy to small enough scales that it can dissipate through heating the plasma. However, the detailed properties of the dissipation process, as well as those of the cascade itself, are still under investigation. Extensive modeling in past years has shown the MHD models of the solar wind do not well constrain the physics of these processes, in that most plausible fluid energy inputs to the coronal plasma will yield a hot, fast wind independent of the details of the heating mechanism.

Thus, it is necessary to incorporate kinetic features of the wind and seek to obtain these features in a kinetic model. A typical fast solar wind plasma exhibits a number of specific kinetic attributes. The ions in the plasma are hotter than the electrons. The proton core is hotter perpendicular to the large-scale magnetic field than parallel to it. Heavy ions are hotter than protons with temperatures matching or exceeding mass-proportionality and they are often seen to flow faster than the protons. Frequently, the plasma also includes secondary beams pointing along the field away from the Sun. Within the context of a turbulent cascade, the kinetic dissipation mechanism, in conjunction with the effects of an expanding flow, should be capable of reproducing these properties.

In this paper, we report the first steps in modeling a specific coronal hole heating mechanism, the dissipation of a critically-balanced turbulent spectrum of kinetic Alfvén waves (KAWs). Observations and simulations of collisionless plasma turbulence find that the fluctuations are akin to randomly-phased fragments of oblique KAWs in that the relations between the fluctuating particles and fields have the same behavior as they would in the waves. Although KAWs were long thought to be dissipated only by Landau damping, Isenberg & Vasquez [1] showed that a cyclotron-resonant interaction was also possible. While the Landau interaction provides particle energization parallel to the magnetic field, the cyclotron interaction can heat ions in the perpendicular directions, in agreement with the coronal hole requirements.

The ion distributions in the near-Sun corona are shaped by the combined actions of the kinetic heating mechanism and the global forces affecting the particles. In order to test the ability of this KAW heating mechanism to reproduce observed fast solar wind distributions, the heating must be incorporated into a kinetic model of coronal hole flow which includes these forces. Here, we take the guiding-center model of Isenberg & Vasquez [2, 3], which describes the radial evolution of collisionless coronal hole protons as they stream away from the Sun under the influence of gravity, ambipolar electric field, ponderomotive force of the large-scale Alfvén waves, and the mirror-force focusing in an expanding flux tube.

## MODEL DESCRIPTION

Apart from the wave-particle interaction providing the heating, the kinetic guiding-center model used here is identical to the model described in Isenberg & Vasquez [3]. The equation for the collisionless steady-state gyrotropic proton distribution function, $f(r, v_\parallel, v_\perp)$, within a radially-directed flux tube, expanding with an area function $A(r)$, is given by

$$(U+v_\parallel)\frac{\partial f}{\partial r} + \left\{-\frac{GM_s}{r^2} + \frac{q}{m}E(r) - (U+v_\parallel)\frac{dU}{dr} - \frac{1}{8\pi\rho}\frac{d<\delta B^2>}{dr}\right\}\frac{\partial f}{\partial v_\parallel}$$

$$+ \frac{v_\perp}{2}\frac{d\ln A}{dr}\left[v_\perp \frac{\partial f}{\partial v_\parallel} - (U+v_\parallel)\frac{\partial f}{\partial v_\perp}\right] = \left(\frac{\partial f}{\partial t}\right)_{w-p/turb} \quad (1)$$

This equation is written in the co-moving frame, so the bulk speed is $U(r)$ and $<v_\parallel f> = 0$, where $<...>$ denotes an average over all $v_\parallel$. The terms on the left-hand side of (1) correspond to advection of the plasma in the radial direction, the radial effects of gravity, ambipolar electric field, inertial force in the accelerating reference frame, ponderomotive force due to the large-scale Alfvén waves, and the mirror force of the expanding magnetic field.

The right-hand side represents the wave-particle heating, in this paper due to cyclotron and Landau damping of KAWs, rather than the ion-cyclotron heating analyzed in [2, 3]. This interaction is modeled as a diffusion in velocity space, according to the quasilinear expressions in [1].

The turbulent spectrum of KAWs is taken to be

$$E_B(\mathbf{k}) = C_{KAW}(r)\frac{(\lambda k_\perp)^{-10/3} + (\lambda k_\perp)^{-2.13}}{1+(\lambda k_\perp)^2} S\left[k_{\parallel cb} - |k_\parallel|\right] \quad (2)$$

where the spectral extent of the parallel wavenumber is limited by the critical balance criterion

$$\lambda k_{\parallel cb} = (\lambda k_i)^{1/3}\frac{(\lambda k_\perp)^{2/3} + (\lambda k_\perp)^{7/3}}{1+(\lambda k_\perp)^2} \quad (3)$$

controlled by $S[...]$, the step function in (2). This expression yields the 1D power laws in an inertial range ($\sim k_\perp^{-5/3}$ for $k_\perp\lambda << 1$) and a dissipation range ($\sim k_\perp^{-2.8}$ for $k_\perp\lambda >> 1$) with a gradual transition around the proton inertial scale, $\lambda = V_A/\Omega_p$. The spectrum is assumed isotropic at a scale $k_i\lambda = 10^{-3}$. These expressions are similar to those in TenBarge & Howes [4] and Isenberg & Vasquez [1], but take into account the geometric dependence of $k_\perp$ in cylindrical coordinates.

The diffusion coefficients $D_{i,j}(v_\parallel, v_\perp)$ are formed by integrating the product of the spectral intensity and the quasilinear resonant fields over $k_\perp$, using the two-fluid dispersion relations of Zhao et al. [5] for KAWs with equal proton and electron temperatures at $\beta = 0.02$. We do not reproduce the lengthy expressions here, but refer to [1] for these details.

As in Isenberg & Vasquez [2, 3, 6], the turbulent fluctuation intensity, $C_{KAW}$, is given a radial profile consistent with the coronal hole model of Cranmer & van Ballegooijen [7]. The computation here is not fully self-consistent in that the spectral shape of the turbulence and the dispersion parameters governing the KAWs are assumed to not change with radial distance from the Sun. However, the flow speed does evolve self-consistently as that value which corresponds to $<v_\parallel f> = 0$ for each radial position.

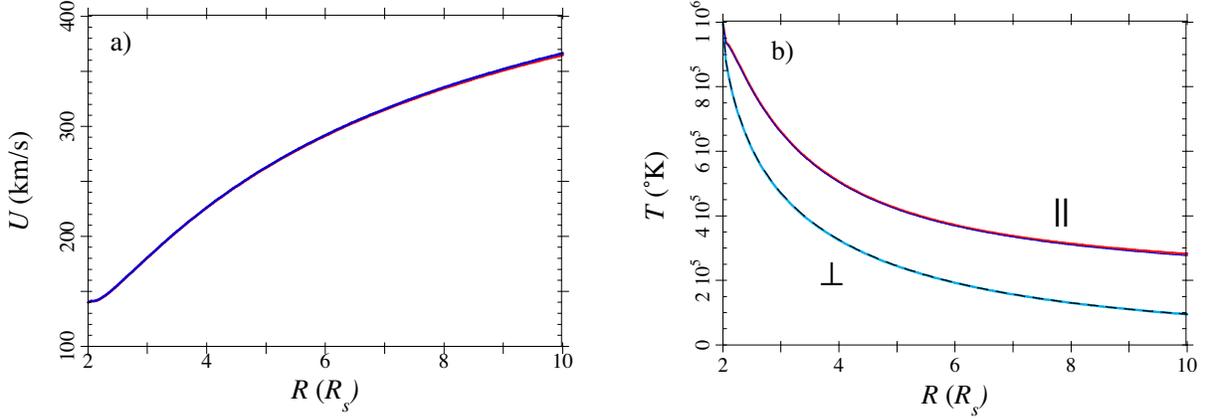

**FIGURE 1.** Bulk parameter results for balanced and imbalanced cases. a) Self-consistently determined flow speeds. b) Parallel and perpendicular proton temperatures, parallel temperatures shown in the upper curves. Results are shown for both cases, but the curves in these plots are almost indistinguishable.

We obtain the steady-state solution for $f$ by solving the time-dependent version of (1), adding a term $\partial f/\partial t$ to the left-hand side, under steady conditions. We start with an isotropic Maxwellian distribution at the inner boundary of $r = 2\ R_s$, with density $n = 2.1\times10^5$ cm$^{-3}$, temperature $T = 10^6$ deg, and Alfvén speed $V_A = 3000$ km/s. The flow speed there is $U\,(2\ R_s) = 140$ km/s and the area function of the flux tube is $A\,(r) = 5\ r^6/[16\ (r^4 + 4)]$ yielding a super-radial expansion factor of 5. The distribution and system parameters are held fixed at this inner boundary, and we set a flow-through condition at the outer boundary of $r = 10\ R_s$. In each time step, the bulk flow at each radial point is re-evaluated according to the new distribution. The computation is run until the solution for $f$ converges to one exhibiting a number flux which is approximately constant in $r$.

Under these conditions, we ran two cases, of balanced and imbalanced turbulence. In the balanced case, the KAW spectrum was taken to be symmetric in $k_{\parallel}$. In the imbalanced case, the outward ($k_{\parallel} > 0$) to inward ($k_{\parallel} < 0$) intensities were set at a ratio of 9:1, while the total intensity was not changed.

## RESULTS

The time-dependent computations were run until the number fluxes between $3 - 10\ R_s$ were conserved to within 0.5% for the balanced case and 1.2% for the imbalanced case. In Fig. 1, we show a) the self-consistently computed flow speeds, and b) the parallel and perpendicular proton temperatures that are produced. The figures show these bulk parameter values for both the balanced and imbalanced cases, but the results yield almost indistinguishable curves on the scale of these plots.

For the turbulent intensities chosen here, we see that the resulting flow speeds are only moderate, while the KAW heating is very weak, falling well short of the requirements for a fast solar wind. Rather than an increase in the perpendicular temperature with increasing radial position, as found in earlier kinetic models [2,3], both perpendicular and parallel temperatures decrease.

The kinetic proton distributions are shown in Fig. 2, for $r = 5.5\ R_s$, and in Fig. 3, for $r = 9.5\ R_s$ (well inside the flow-through boundary at $10\ R_s$). The balanced and imbalanced cases can be distinguished kinetically, but primarily in the extreme halos of the distributions where the imbalanced turbulence is seen to heat the sunward protons more than in the balanced case. While the perpendicular heating needed to reach fast wind speeds is not at all evident, we note that the distributions spread in the outward parallel direction due to the Landau resonance with the KAWs. We anticipate that this effect, when coupled to an effective perpendicular heating mechanism, will contribute to the production of the secondary proton beams that are often observed in the fast solar wind.

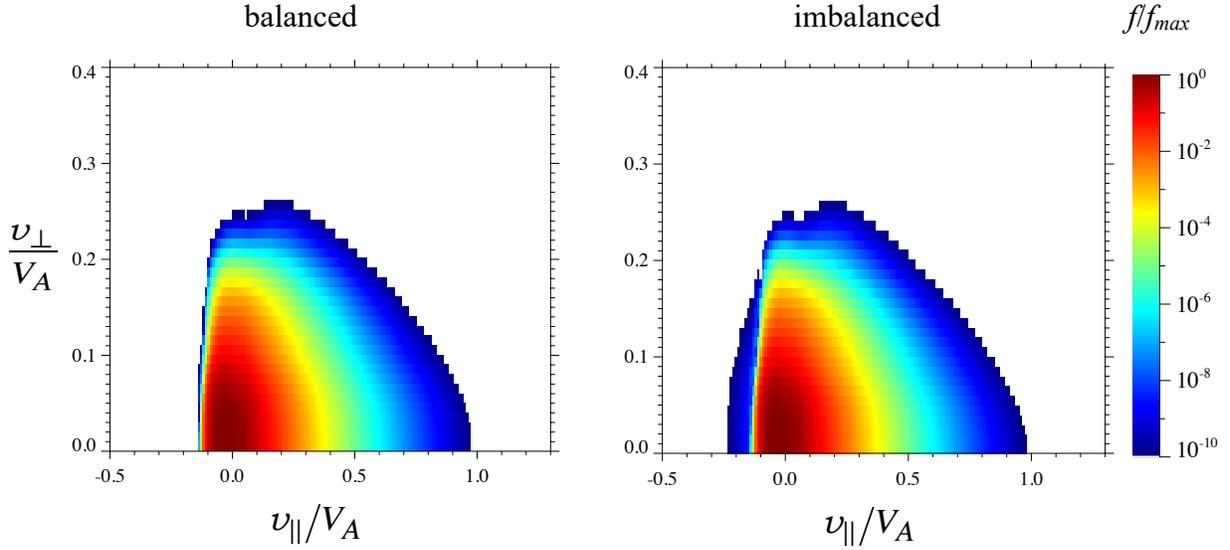

**FIGURE 2.** Proton distribution functions at $r = 5.5\ R_s$, normalized to the maximum value of the phase-space density.

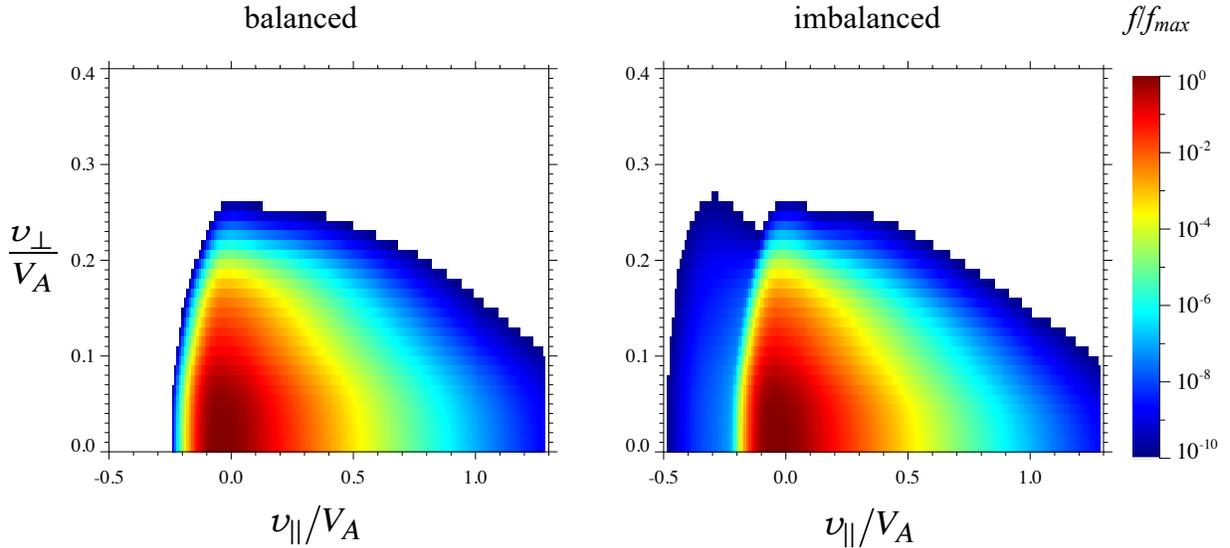

**FIGURE 3.** Proton distribution functions at $r = 9.5\ R_s$, normalized to the maximum value of the phase-space density.

## CONCLUSIONS

We conclude that the dissipation of turbulent KAWs, as modeled here, is not sufficient to generate a fast solar wind. It appears, in fact, that the proton acceleration that does result is caused by the ponderomotive force of the large-scale Alfvén waves in the model. It is possible that modifying some of the model parameters could give greater heating and acceleration, but we suspect that substantial, and less plausible, changes would be required.

A more fruitful direction would be to modify the model along the lines of the recent helicity barrier hypothesis [8, 9]. Simulations and observations have suggested that strongly imbalanced turbulence would not be allowed to proceed to the dissipation scales at the rate implied by the Kolmogorov or Kraichnan cascade of inertial-range intensities. Rather, the intensities in the majority component are reduced at the high-$k$ end of the inertial range, through a transition range, to the level of the minority component before cascading into the dissipation range. The turbulent power removed from the inertial-range majority component by this transition is transformed into quasi-parallel ion-cyclotron fluctuations that then dissipate through the standard cyclotron resonance. We will extend the model of this paper to explore the predictions of coronal hole acceleration and heating under such a helicity barrier scenario.

## ACKNOWLEDGEMENTS

The authors thank Trevor Bowen, Jono Squire, Jason TenBarge, and Daniel Verscharen for valuable conversations. P.A.I. and B.J.V. are supported by NSF grant AGS2005982. Additionally, B.J.V. is supported by NASA grant 80NSSC21K1674.

## REFERENCES

1. P. A. Isenberg and B. J. Vasquez, Astrophys. J. **887**, 63 (2019).
2. P. A. Isenberg and B. J. Vasquez, Astrophys. J. **731**, 88 (2011).
3. P. A. Isenberg and B. J. Vasquez, Astrophys. J. **808**, 119 (2015).
4. J. M. TenBarge and G. G. Howes, Phys. Plasmas **19**, 055901 (2012).
5. J. S. Zhao, Y. Voitenko, M. Y. Yu, J. Y. Lu and D. J. Wu, Astrophys. J. **793**, 107 (2014).
6. P. A. Isenberg and B. J. Vasquez, Astrophys. J. **696**, 591 (2009).
7. S. R. Cranmer and A. A. van Ballegooijen, Astrophys. J. Suppl. **156**, 265 (2005).
8. R. Meyrand, J. Squire, A. A. Schekochihin and W. Dorland, J. Plasma Phys. **87**, 535870301 (2021).
9. J. Squire, R. Meyrand, M. W. Kunz, L. Arzamasskiy, A. A. Schekochihin and E. Quataert, Nature Astron. **6**, 715 (2022).